\documentclass[aps,showpacs,twocolumn,prl]{revtex4}
\usepackage{graphicx}

\begin{document}

\newcommand{\pderiv}[2]{\frac{\partial #1}{\partial #2}}
\newcommand{\deriv}[2]{\frac{d #1}{d #2}}
\newcommand{\eq}[1]{Eq.~(\ref{#1})}  
\newcommand{\infint}{\int \limits_{-\infty}^{\infty}}

\title{Nonlinear Schroedinger Equation
in the Presence of Uniform Acceleration}

\vskip \baselineskip

\author{A.R. Plastino$^{1,2}$}\thanks{Corresponding Author: arplastino@ugr.es}
\author{C. Tsallis$^{3,4}$}

\address{
$^1$Instituto de Fisica Teorica y Computacional Carlos I,
Universidad de Granada, Granada, Spain \\
$^2$National University La Plata-CREG, Casilla de Correos 727, 1900 La Plata, Argentina \\
$^{3}$Centro Brasileiro de Pesquisas F\'{\i}sicas and 
National Institute of Science and Technology for Complex Systems, 
Rua Xavier Sigaud 150, 
22290-180  Rio de Janeiro-RJ, Brazil \\
$^{4}$Santa Fe Institute, 1399 Hyde Park Road, Santa Fe, New Mexico 87501,
USA}

\date{\today}

\newpage

\begin{abstract}
We consider a recently proposed nonlinear Schroedinger equation exhibiting soliton-like 
solutions of the power-law form $e_q^{i(kx-wt)}$, involving the $q$-exponential function 
which naturally emerges within nonextensive thermostatistics [$e_q^z \equiv [1+(1-q)z]^{1/(1-q)}$, 
with $e_1^z=e^z$]. Since these basic solutions behave like free particles, obeying $p=\hbar k$, 
$E=\hbar \omega$ and $E=p^2/2m$ ($1 \le q<2$), it is relevant to investigate how they change 
under the effect of uniform acceleration, thus providing the first steps towards the 
application of the aforementioned nonlinear equation to the study of physical scenarios 
beyond free particle dynamics. We investigate first the behaviour of the power-law solutions 
under Galilean transformation and discuss the ensuing Doppler-like effects.
 We consider then constant acceleration, obtaining new solutions that can be equivalently 
regarded as describing a free particle viewed from an uniformly accelerated reference frame
(with acceleration $a$) or a particle moving under a constant force $-ma$. The latter 
interpretation naturally leads to the evolution equation 
$i\hbar \frac{\partial}{\partial t}\left( \frac{\Phi}{\Phi_0} \right) \,\, = \,\,
- \frac{1}{2-q}\frac{\hbar^2}{2m} \frac{\partial^2}{\partial x^2}
\left[\left( \frac{\Phi}{\Phi_0} \right)^{2-q}\right] +
V(x)\left( \frac{\Phi}{\Phi_0} \right)^{q}$ with $V(x)=max$. Remarkably enough, 
the potential $V$ couples to $\Phi^q$, instead of coupling to $\Phi$, 
as happens in the familiar linear case ($q=1$).

\vskip \baselineskip

\end{abstract}
\maketitle

The importance of symmetries in physics can hardly be overestimated.
A proper understanding of the relevant symmetries exhibited by a physical 
phenomenon encapsulates some of its most essential features, and usually 
leads to the most elegant and compact mathematical formulation of the
associated physical laws. Many important symmetries are directly 
related to changes in the reference frame employed to describe physical 
systems. Such is the case of the dynamical symmetries associated with the
Galilean transformation between inertial frames, playing a central role 
in Newtonian mechanics and non-relativistic quantum mechanics; the 
Lorentz transformation plays an analogous role in special relativity, 
both at the classical and at the quantum levels.
We can also mention the invariance of statistical mechanics and thermodynamics 
under uniform translation of the energy spectrum, which leads to the full 
freedom of choosing the zero level of energies in any physical system
(the above list of examples is, of course, far from complete).
A remarkable feature of symmetry or invariance principles is that 
they constitute powerful heuristic tools to extend knowledge 
about particular or restricted instances of the 
behaviour of physical systems to more general situations or scenarios.
The most spectacular example of this approach was probably Einstein's 
use of his Principle of Equivalence in the development of General Relativity.

Symmetries with important physical implications are also observed in 
pure mathematics. We may focus, for example, on the Central Limit Theorem: 
the stability of Gaussians in the space of distributions with regard to adding 
(into the system) a finite or even infinite number of nearly independent 
random variables with finite variance is at the heart of their robustness, 
and therefore of their ubiquity in nature.
The expansibility property of an entropic functional (that is, its invariance
when adding new microscopic configurations with zero probability) constitutes
another relevant example. It plays a special role within the (Shannon-Khinchine) 
set of axioms that uniquely determine the mathematical form of the Boltzmann-Gibbs 
(BG) additive entropy, basis of BG statistical mechanics.

Symmetry considerations arise naturally in the study of both linear and nonlinear
dynamics. In the latter case, however, the analysis of the relevant symmetries is
usually much more difficult. The physically relevant nonlinear evolution equations
are highly diverse and describe wide classes of physical systems: see for 
instance \cite{sulem,frankbook,scott0507,polyanin}. 
Among the most studied nonlinear differential equations we have nonlinear versions of the  
Schroedinger~\cite{sulem,NobreMonteiroTsallis2011,NobreMonteiroTsallis2012} 
and Fokker-Planck~\cite{frankbook,CearaRio2010,Mauricio2012} ones. 
Here we focus primarily on the soliton-like solutions of the 
recently proposed nonlinear Schroedinger equation \cite{NobreMonteiroTsallis2011}
inspired by nonextensive statistical mechanics and the associated 
nonadditive entropies \cite{tsallis88,tsallisbook,B09}. The nonlinear evolution equation
advanced in \cite{NobreMonteiroTsallis2011} (and the above mentioned soliton-like solution)
are implicitly assumed to hold with repect to an inertial reference frame. Our 
main aim in the present work is to investigate the form of these solutions when 
described with respect to a uniformly accelerated reference frame. We 
consider first the behavior of the soliton-like solutions under
Galilean transformations connecting inertial frames. We then tackle
the case of accelerated frames.

The aforementioned soliton-like solutions are referred to as $q$-plane waves 
and may be relevant in diverse areas of physics, including nonlinear optics, 
superconductivity, plasma physics, and dark matter \cite{galgani,NobreMonteiroTsallis2012}. 
The theory within which $q$-plane waves emerged generalizes the BG 
entropy and statistical mechanics, through the introduction of an index $q$ 
($q \rightarrow 1$ recovers the BG case). Along this line, considerable progress 
was achieved, leading to generalized functions, distributions, various equations 
of physics, and new forms of the Central Limit Theorem~\cite{CLT}.
In particular, the $q$-Gaussian distribution, which generalizes 
the standard Gaussian, appears naturally by extremizing 
the $q$-entropy~\cite{tsallis88},  
or from the solution of the corresponding nonlinear Fokker-Planck
equation~\cite{plastino95},  and has  
been successfully applied to the analysis of 
recent experimental results in various fields~\cite{tsallisbook}. 
Among others, we may mention:
(i) The velocities of cold atoms in dissipative optical 
lattices~\cite{douglas06};  
(ii) The velocities of particles in quasi-two dimensional dusty 
plasma~\cite{liugoreeprl08};
(iii) Single ions in radio frequency traps interacting with a classical
buffer gas~\cite{devoe}; 
(iv) The relaxation curves of RKKY spin glasses, like CuMn and
AuFe~\cite{pickup}; 
(v) Transverse momenta distributions at LHC
experiments~\cite{CMS}.

Herein we discuss the following  $q$-generalized Schroedinger equation 
for a $d$-dimensional free particle of mass $m$ \cite{NobreMonteiroTsallis2011}:
\begin{equation}
\label{eq:schreq}
i \hbar {\partial \over \partial t} 
\left[ \frac{\Phi(\vec{x},t)}{\Phi_{0}} \right] 
= - {1 \over 2-q} \ \frac{\hbar ^{2}}{2m}
\nabla^{2} \left[ \frac{\Phi(\vec{x},t)}{\Phi_{0}} \right]^{2-q}\;(q \ge 1) \,,
\end{equation}
where $\Phi_{0}$ guarantees the correct physical dimensionalities for all terms 
(this scaling becomes 
irrelevant only for the linear equation, i.e., $q=1$). 
Its solutions are expressed in terms of
the $q$-exponential function $\exp_{q}(u)$ which, for a pure
imaginary $iu$, is defined as the principal value of
\begin{equation}
\label{eq:compqexp}
\exp_{q}(iu) = \left[ 1 + (1-q)iu \, \right]^{\frac{1}{1-q}}; 
\,\exp_{1}(iu) \equiv \exp(iu). 
\end{equation}
The above function satisfies~\cite{borges98},
\begin{eqnarray}
\label{eq:propcompqexp1}
\exp_{q}(\pm iu) &=& \cos_{q}(u) \pm i \sin_{q}(u)~, \cr
\cos_{q}(u) &=& \rho_{q}(u)
\cos \left\{ {1 \over q-1} {\rm arctan}[(q-1)u] \right\}~, \cr
\sin_{q}(u) &=& \rho_{q}(u)
\sin \left\{ {1 \over q-1} {\rm arctan}[(q-1)u] \right\}~, \cr
\rho_{q}(u) &=& \left[1+(1-q)^{2}u^{2} \right]^{1/[2(1-q)]}~,
\end{eqnarray}
\begin{eqnarray}
\label{eq:propcompqexp5}
\exp_{q}(iu)\exp_{q}(-iu) \! & = & 
\! [\rho_{q}(u)]^2  \! = \! \exp_q(-(q-1)u^2), \cr
\exp_{q} (i u_{1})  \exp_{q} (i u_{2})
\! & \ne & \! \exp_{q} \left[ i(u_{1} + u_{2}) \right], \,\, (q\ne1)
\end{eqnarray}  
As a consequence of Eqs.~(\ref{eq:propcompqexp1}-\ref{eq:propcompqexp5}),
a $q$-exponential with a pure imaginary argument, $\exp_{q}(iu)$,
presents an oscillatory behavior with a $u$-dependent amplitude $\rho_{q}(u)$.
The function $\exp_{q}(iu)$ complies with the physically important property 
of square integrability for $1<q<3$, whereas the concomitant
integral diverges in both limits $q \rightarrow 1$ and $q \rightarrow 3$
 and also for $q<1$~\cite{max}.

The $d$-dimensional $q$-plane wave is given by 
\begin{equation}
\label{eq:3dqsolwaveeq}
\Phi(\vec{x},t) = \Phi_{0} \, \exp_{q} \left[ i (\vec{k} \cdot \vec{x}
-\omega t) \right]~, 
\end{equation}
If we take into account that $d\exp_q(z)/dz=[\exp_q(z)]^q$ and
$d^2\exp_q(z)/dz^2=q[\exp_q(z)]^{2q-1}$ we obtain, for the $d$-dimensional
Laplacian, 
\begin{equation}
\label{eq:nabla2qexp}
\nabla^{2} \left(\frac{\Phi}{\Phi_0} \right)= 
- q \left( \sum_{n=1}^{d} k_{n}^{2} \right) \left( \frac{\Phi}{\Phi_0} \right)^{2q-1}.
\end{equation}
Now, inserting the $q$-plane wave ansatz
(\ref{eq:3dqsolwaveeq}) into the NL Schroedinger Eq. (\ref{eq:schreq}), we
verify that the $q$-plane wave is indeed a solution provided that
the frequency $\omega$ and the momentum $k$ satisfy the relation
$\omega = \frac{\hbar k^2}{2m}$. Equivalently, if one makes \cite{NobreMonteiroTsallis2011}, according 
to the celebrated de Broglie and Planck relations, the identifications 
$\vec{k} \rightarrow \vec{p}/\hbar$ and $\omega \rightarrow E/\hbar$, 
one verifies that the $q$-plane wave is a solution of equation 
(\ref{eq:schreq}) with $E=p^{2}/2m$, thus {\it preserving the 
energy spectrum of the free particle for all values of $q$}.

Eq.~(\ref{eq:schreq}) differs from previous 
formulations~\cite{sulem,KLQ98} where new nonlinear terms
 (usually a cubic nonlinearity in the wave function) 
are added to the two existing linear terms.
The main differences between~\eq{eq:schreq} and 
other proposals for NL Schroedinger equations are: 
(i) Instead of adding an extra term in which the nonlinearity is
introduced, we generalize the spatial second-derivative term; 
(ii) The equation, together with the proposed solution, are  
easily extended from one to $d$ dimensions;  
(iii) The corresponding solution of~\eq{eq:schreq} manifests nonlinearity
in both space and time, through a modulation in these two variables,  
which keeps the norm finite for all $(\vec{x},t)$; 
(iv) The well-known energy spectrum of a free particle is preserved 
for all $q$. 
Therefore \eq{eq:schreq}, together with its solution \eq{eq:3dqsolwaveeq}, 
can be considered as candidates for describing interesting types 
of physical phenomena.

 Let us now investigate how the $q$-plane wave solution is transformed under
two basic types of changes in the reference frame. As already mentioned, we consider first a Galilean transformation connecting inertial reference
frames. Then, we analyze the aspect of the $q$-plane wave solutions
when ``viewed'' from an uniformly accelerated reference frame.
We shall consider the one dimensional nonlinear Schroedinger equation. 
The extension to the $d$-dimensional case is straightforward.

We shall assume that the nonlinear Schroedinger equation,
\begin{equation}
\label{eq:schreq_primada}
i \hbar {\partial \over \partial t^{\prime}} 
\left[ \frac{\Psi(x^{\prime},t^{\prime})}{\Psi_{0}} \right] 
= - {1 \over 2-q} \ \frac{\hbar ^{2}}{2m} \frac{\partial^2}{\partial x^{\prime 2}}
\left[ \frac{\Psi(x^{\prime},t^{\prime})}{\Psi_{0}} \right]^{2-q}\;(q \ge 1),
\end{equation}
holds in an inertial reference frame characterized by the spatio-temporal coordinates 
$(x^{\prime}, t^{\prime})$ and that the system under consideration is described by 
a $q$-plane wave solution 
\begin{equation} \label{soluprim}
\Psi(x^{\prime},t^{\prime}) = \Psi_{0} \, \exp_{q} 
\left[ i (k x^{\prime} -\omega t^{\prime}) \right].
\end{equation}
Let us consider now a Galilean transformation
\begin{equation}
t \, = \, t^{\prime}; \,\,\,\,\,\, x \, = \, x^{\prime} - v t^{\prime},
\end{equation}
relating the original inertial frame $(x^{\prime}, t^{\prime})$
with a second inertial frame $(x,t)$ that moves with respect 
to the former one with a uniform velocity $v$. To obtain a
``new'' solution $\Phi(x,t)$ to the nonlinear Schroedinger equation 
expressed in terms of the spatio-temporal coordinates $(x,t)$, 
that corresponds to the ``old'' solution (\ref{soluprim}), one 
may naively just re-express (\ref{soluprim}) in terms of the new 
variables (that is, substitute in (\ref{soluprim}) the old 
variables $(x^{\prime},t^{\prime})$ by their expressions
in terms of the new variables $(x,t)$),
\begin{eqnarray}
\Psi(x^{\prime},t^{\prime})\!\! &\rightarrow & \!\! 
\Phi(x,t)  \!\!  =  \!\!  \Psi(x+vt,t) \cr
&=& \!\!\! \Phi_{0}  \exp_{q} 
\left[i (k (x+vt)\! - \!\omega t) \right]\!, \,\, \Phi_0 \!=\! \Psi_0
\end{eqnarray}
 However, this procedure leads 
to a function of $x$ and $t$ that does not satisfy the nonlinear 
Schroedinger equation (as expressed in terms of the new variables $(x,t)$).
In order to obtain a valid solution it is necessary to add an extra term
to the argument of the $q$-exponential. Indeed, it can be verified after 
some algebra that,
\begin{equation}
\label{eq:phiunivel}
\frac{\Phi}{\Phi_0}\! =\! \left[\!
1 - i (1-q)\left\{
\omega t - k \!\left(\! x \! + \! vt \!\right)\! +
\frac{1}{\hbar} \!\left(\! m v x \! + \!\frac{1}{2} mv^2 t \!\right)\!
\right\}\!\!
\right]^{\frac{1}{1-q}}.
\end{equation}
does satisfy the nonlinear Schroedinger equation. 
The extra term $\frac{1}{\hbar} \!\left(\! m v x \! 
+ \!\frac{1}{2} mv^2 t \!\right)$ admits a clear physical
interpretation. If we recast (\ref{eq:phiunivel})
under the guise,
\begin{equation}
\label{eq:phiunivelo}
\frac{\Phi}{\Phi_0}\! =\! \left[\!
1 - i (1-q)\left\{
\left( \omega - kv  + \frac{mv^2}{2 \hbar} \right) t - 
\left(k  - \frac{m v}{\hbar}  \right) x
\right\}\!\!
\right]^{\frac{1}{1-q}},
\end{equation}
it is plain that (\ref{eq:phiunivelo}) has the form
of a $q$-plane wave with frequency $\tilde \omega$
and wave number $\tilde k$ respectively given by
\begin{equation} \label{omegakatran}
{\tilde \omega} \, = \, \omega - kv  + \frac{mv^2}{2 \hbar}; \,\,\,\,\,\,\,\,
{\tilde k} \, = \, k  - \frac{m v}{\hbar}. 
\end{equation}
Now, as shown in \cite{NobreMonteiroTsallis2011}, 
the $q$-plane wave solutions
to the nonlinear Schroedinger equation are compatible 
with the Planck and de Broglie relations connecting respectively frequency
and wave number with energy and momentum:
$E = \hbar \omega$ and $p = \hbar k$. Combining equations
(\ref{omegakatran}) with the Planck and de Broglie relations we
obtain ${\tilde E}= E - pv + \frac{mv^2}{2}$ and
${\tilde p} = p - mv$, which are the correct Galilean 
transformations for the kinetic energy and momentum of 
a particle of mass $m$ obeying the (nonrelativistic)
energy-momentum relation $E=p^2/2m$. These considerations reinforce
the validity of the Planck and de Broglie relations for the $q$-plane wave
solutions of \eq{eq:schreq}.

Taking the limit $q\to 1$ of the transformed solution 
(\ref{eq:phiunivel}), on sees that the relation between 
the original solution $\Psi(x^{\prime}, t^{\prime})$ and 
the transformed one $\Phi(x,t)$ becomes,
\begin{eqnarray}
\Psi(x^{\prime}, t^{\prime}) \!\!&\rightarrow & \!\!\Phi(x,t) \cr
\!\!& = &\!\! \exp\left[\!-\frac{i}{\hbar}\! 
\left(\!  \frac{mv^2}{2} t + mvx \! \right) 
\!\right]\!\! \Psi(x+vt,t),
\end{eqnarray}
thus recovering the transformation rule corresponding to the 
linear Schroedinger equation \cite{P93}.

 Let us now consider a uniformly accelerated reference frame.
The corresponding spatio-temporal coordinates $(x,t)$ are
\begin{equation}
t \, = \, t^{\prime}; \,\,\,\,\,\,\,\,
x \, = \, x^{\prime} - \frac{1}{2} a t^{\prime 2}
\, = \, x^{\prime} - \frac{1}{2} \frac{F}{m} t^{\prime 2} \,,
\end{equation}
where $(x^{\prime},t^{\prime})$ are the variables
associated with an inertial frame, $a$ is the constant acceleration
of reference frame $(x,t)$, and $a = \frac{F}{m}$. As
in the previous discussion, we assume that the nonlinear
Schroedinger equation (\ref{eq:schreq_primada}) holds in the inertial
frame $(x^{\prime},t^{\prime})$, and also that in this frame
our system is described by the $q$-plane wave solution (\ref{soluprim}).
Again, simply re-writting the $q$-plane wave solution (\ref{soluprim})
in terms of the new variables $(x,t)$ does not yield a solution
of the nonlinear Schroedinger equation. As in the above
Galilean transformation case, new terms are needed
in the argument of the $q$-exponential to obtain a valid solution.
Let us consider the ansatz,
\begin{equation} \label{accelequwave}
\label{eq:phiacce}
\frac{\Phi}{\Phi_0}\! =\! \left[\!
1 - i (1-q)\left\{
\omega t - k \!\left(\! x \! + \!\frac{Ft^2 }{2m} \!\right)\! +
\frac{F}{\hbar} \!\left(\! xt \! + \!\frac{Ft^3 }{6m} \!\right)\!
\right\}\!\!
\right]^{\frac{1}{1-q}}.
\end{equation}
Inserting (\ref{accelequwave}) in the right and  the left hand sides of the 
nonlinear Schroedinger equation yields
\begin{equation} \label{leftaccequwa}
\label{eq:phiacce2}
i\hbar \frac{\partial}{\partial t}\left( \frac{\Phi}{\Phi_0} \right) \!\! = \!\!
\left[ \hbar \omega - \frac{\hbar k F t}{m} + F x +
\frac{F^2t^2 }{2m} \right] \!\!
\left( \frac{\Phi}{\Phi_0} \right)^{\!q}\!\!,
\end{equation}
and
\begin{equation} \label{rightaccequwa}
\label{eq:phiacce3}
- \frac{1}{2-q}\frac{\hbar^2}{2m} \frac{\partial^2}{\partial x^2}\!
\left[\!\!\left( \frac{\Phi}{\Phi_0} \right)^{2-q}\right]
\!\! =\!\! \left[\! \frac{\hbar^2 k^2}{2m} - \frac{\hbar k F t}{m} +
\frac{F^2t^2 }{2m} \right]\!\!
\left(\! \frac{\Phi}{\Phi_0} \!\right)^{\!q}\!\!.
\end{equation}
Comparing (\ref{leftaccequwa}) with (\ref{rightaccequwa})
one verifies that the ansatz (\ref{accelequwave}) satisfies
the nonlinear equation,
\begin{equation} 
\label{constantforce}
i\hbar \frac{\partial}{\partial t}\left( \frac{\Phi}{\Phi_0} \right) \!\! = \!\!
- \frac{1}{2-q}\frac{\hbar^2}{2m} \frac{\partial^2}{\partial x^2}
\left[\left( \frac{\Phi}{\Phi_0} \right)^{\!\!  2-q}\right] +
V(x)\left( \frac{\Phi}{\Phi_0} \right)^{\!q}\!\!,
\end{equation}
where $V(x) = F x$. The nonlinear Eq. (\ref{constantforce})
can be interpreted as describing the motion of a particle of mass $m$
under a constant force $-F$ (with the associated potential function
$V = Fx$). This is consistent with the well-known fact that the
behavior of a free particle with respect to a uniformly 
accelerated reference frame is equivalent to the the behavour of a
particle in an inertial reference frame moving under the effect 
of a constant force. In the limit $F \to 0$, Eq. 
(\ref{constantforce}) reduces to the nonlinear Schroedinger equation
for a free particle introduced in \cite{NobreMonteiroTsallis2011},
and solution (\ref{accelequwave}) reduces to the corresponding
$q$-plane wave solution. Also, $q \to 1$ in Eq. (\ref{constantforce}) corresponds to the standard linear
Schroedinger equation for a particle of mass $m$ moving under 
a constant force $-F$. An interesting feature of equation 
(\ref{constantforce}) is that the potential $V$ couples 
to $\Phi^q$, instead of coupling to $\Phi$, as happens in 
the standard linear case ($q=1$). Consistently,
the $q$-plane wave $\Phi(x,t)=\Phi_{0} \, \exp_{q} 
\left[ i (k x -\omega t) \right]$ is not only a solution of the
free-particle nonlinear Schroedinger equation (when $\hbar \omega = \frac{\hbar^2 k^2}{2m}$),
but also of the nonlinear equation
\begin{equation}
\label{conpote}
i\hbar \frac{\partial}{\partial t}\left( \frac{\Phi}{\Phi_0} \right) \!\! = \!\!
- \frac{1}{2-q}\frac{\hbar^2}{2m} \frac{\partial^2}{\partial x^2}
\left[\left( \frac{\Phi}{\Phi_0} \right)^{\!\!  2-q}\right] +
V_0\left( \frac{\Phi}{\Phi_0} \right)^{\!q}\!\!,
\end{equation}
with a constant potential $V_0$, provided that
$ \hbar \omega = \frac{\hbar^2 k^2}{2m} + V_0$,
which, using the Planck and de Broglie relations, 
becomes $E = \frac{p^2}{2m} + V_0$, as expected.

Considering now the limit $q\to 1$ of the transformed solution 
(\ref{accelequwave}), we verify that the original solution 
$\Psi(x^{\prime}, t^{\prime})$ and the transformed one $\Phi(x,t)$ 
are linked through
\begin{eqnarray}
\Psi(x^{\prime}, t^{\prime}) \!\!&\rightarrow & \!\!\Phi(x,t) \cr
\!\!& = &\!\! \exp \!\left[\!
-\frac{i}{\hbar} \!\left(\! Fxt \! + \!\frac{F^2t^3 }{6m} \!\right)
\!\right]\!\! \Psi \!\left(\! x \! + \!\frac{Ft^2 }{2m},t\!\right) \!,
\end{eqnarray}
thus recovering the transformation rule associated with the 
linear Schroedinger equation \cite{P93}.

We have investigated the effect of uniform acceleration on
$q$-plane waves, soliton-like solutions of a recently proposed
non-linear Schroedinger equation associated with non-extensive
thermostatistics. We first studied the behaviour
of these solutions under Galilean transformations relating different
inertial frames and obtained the transformation rule satisfied
by the $q$-plane waves. This rule turns out to be fully consistent 
with the de Broglie and Planck relations, thus providing further support to 
the validity of these relations for the $q$-plane wave solutions.
Then we derived the transformation law yielding new solutions
corresponding to the aforementioned $q$-plane waves when ``viewed''
from uniformly accelerated frames. In the limit $q\to 1$ the transformation
laws advanced here reduce to those associated with time dependent
solutions of the standard, linear Schroedinger equation. The accelerated 
$q$-plane wave solutions investigated here admit two possible
interpretations: they can be viewed as describing a free particle as
``seen'' from a uniformly accelerated frame or, alternatively,
as describing a particle moving under the effect of a constant force.
In fact, the non-linear Schroedinger equation governing these
solutions (when expressed in terms of the accelerated frame's
coordinates) incorporates a new term involving a potential function $V(x)$.
This equation indicates that, within the present generalization
 of Schroedinger equation, the potential $V(x)$ ``couples'' to a
power of the wave function, $\Phi^q$, instead of coupling just to 
the wave function $\Phi$, as happens with the linear Schroedinger 
equation. This result opens the door for the study of a variety of 
simple potentials, which should result in physical applications.

\noindent
Partial financial support from CNPq and FAPERJ (Brazilian agencies),
and from the Projects FQM-2445 and FQM-207 of the Junta de Andalucia
is acknowledged.


\begin{thebibliography}{40}

\bibitem{sulem}
C. Sulem and P.-L. Sulem, {\it The Nonlinear Schr\"odinger 
Equation: Self-Focusing and Wave Collapse} (Springer, New York, 1999).

\bibitem{frankbook}
T.D. Frank, {\it Nonlinear Fokker-Planck Equations: Fundamentals and
Applications} (Springer, Berlin, 2005).

\bibitem{scott0507}
A.C. Scott,
{\it The Nonlinear Universe} (Springer, Berlin, 2007).

\bibitem{polyanin}
A.D. Polyanin and V.F. Zaitsev, {\it Handbook of Nonlinear Partial 
Differential Equations} (Chapman and Hall / CRC, Boca Raton, 2004).

\bibitem{NobreMonteiroTsallis2011}
F.D. Nobre, M.A. Rego-Monteiro and C. Tsallis, Phys. Rev. Lett. {\bf 106}, 140601 (2011).

\bibitem{NobreMonteiroTsallis2012}F.D. Nobre, M.A. Rego-Monteiro and C. Tsallis, EPL {\bf 97}, 41001 (2012).

\bibitem{CearaRio2010}J. S. Andrade Jr., G.F.T. da Silva, A.A. Moreira, 
F.D. Nobre and E.M.F. Curado, Phys. Rev. Lett. {\bf 105}, 260601 (2010);
Y. Levin and R. Pakter, Phys. Rev. Lett. {\bf 107},   088901 (2011); 
J. S. Andrade Jr., G.F.T. da Silva, A.A. Moreira, F.D. Nobre and 
E.M.F. Curado, Phys. Rev. Lett. {\bf 107}, 088902 (2011).

\bibitem{Mauricio2012}M.S. Ribeiro, F.D. Nobre and E.M.F. Curado, 
Phys. Rev. E {\bf 85}, 021146 (2012).

\bibitem{tsallis88}
C. Tsallis, J. Stat. Phys. {\bf 52}, 479 (1988).

\bibitem{tsallisbook}
C. Tsallis, {\it Introduction to Nonextensive Statistical Mechanics} 
(Springer, New York, 2009).

\bibitem{B09} C. Beck,  Contemporary Physics {\bf 50}, 495 (2009).


\bibitem{galgani}A. Carati, S.L. Cacciatori and L. Galgani, EPL {\bf 83}, 
59002 (2008).

\bibitem{CLT}S. Umarov, C. Tsallis and S. Steinberg, Milan J. Math. {\bf 76}, 307 (2008); 
S. Umarov, C. Tsallis, M. Gell-Mann and S. Steinberg, J. Math. Phys. {\bf 51}, 033502 (2010). 

\bibitem{plastino95}
A.R. Plastino and A. Plastino, Physica A {\bf 222}, 347 (1995); 
C. Tsallis and D.~J. Bukman, Phys. Rev. E {\bf 54}, R2197 (1996).

\bibitem{douglas06}
P. Douglas, S. Bergamini, and F. Renzoni, Phys. Rev. Lett. {\bf 96}, 
110601 (2006). 

\bibitem{liugoreeprl08}
B. Liu and J. Goree, Phys. Rev. Lett. {\bf 100}, 055003 (2008). 

\bibitem{devoe}
R.G. DeVoe, Phys. Rev. Lett. {\bf 102}, 063001 (2009). 

\bibitem{pickup}
R.~M. Pickup, R. Cywinski, C. Pappas, B. Farago, and P. Fouquet, 
Phys. Rev. Lett. {\bf 102}, 097202 (2009). 

\bibitem{CMS}V. Khachatryan et al. (CMS Collaboration), Phys. Rev. 
Lett. {\bf 105}, 022002 (2010).

\bibitem{borges98}
E.P. Borges, J. Phys. A {\bf 31}, 5281 (1998).

\bibitem{max}
M. Jauregui and C. Tsallis, J. Math. Phys. {\bf 51}, 063304 (2010); 
A. Chevreuil, A. Plastino and C. Vignat, J. Math. Phys. {\bf 51}, 
093502 (2010).

\bibitem{KLQ98} G. Kaniadakis, A. Lavagno, and P. Quarati
Phys. Rev. E 57, 1395 (1998).

\bibitem{P93} A. Peres, {\it Quantum Theory: Concepts and Methods} (Kluwer, Dordrecht, 1993).


\end{thebibliography}
\end{document}